\documentclass{epl}

\title{Magnetic oscillations in a two-dimensional network
of compensated electron and hole orbits}
\author{A. Audouard\inst{1}\thanks{E-mail: \email{audouard@lncmp.org}}, D. Vignolles\inst{1}, E.
Haanappel\inst{1}, I. Sheikin\inst{2},
R.~B.~Lyubovski\u{i}\inst{3} and R. N. Lyubovskaya\inst{3}}

\institute{
  \inst{1} Laboratoire National des Champs Magn\'{e}tiques
Puls\'{e}s (UMR CNRS-UPS-INSA 5147), 143 avenue de Rangueil, 31432
Toulouse cedex 4, France\\
  \inst{2} Grenoble High Magnetic Field
Laboratory, CNRS, BP166, 38042 Grenoble Cedex 9, France\\
  \inst{3} Institute of Problems of Chemical Physics, Russian
Academy of Sciences, Chernogolovka, Moscow oblast, 142432 Russia
 }

\pacs{71.18.+y}{Fermi surface: calculations and measurements;
effective mass, g factor} \pacs{72.15.Gd}{Galvanomagnetic and
other magnetotransport effects} \pacs{71.20.Rv}{Polymers and
organic compounds}

\shortauthor{D. Vignolles \etal} \shorttitle{Magnetic oscillations
in a two-dimensional network}

\begin{document}

\maketitle

\begin{abstract}
The Fermi surface of the quasi-two dimensional (2D) organic metal
(ET)$_8$Hg$_4$Cl$_{12}$ (C$_6$H$_5$Br)$_2$ can be regarded as a 2D
network of compensated electron and hole orbits coupled by
magnetic breakthrough. Simultaneous measurements of the interlayer
magnetoresistance and magnetic torque have been performed for
various directions of the magnetic field up to 28 T in the
temperature range from 0.36 K to 4.2 K. Magnetoresistance and de
Haas-van Alphen (dHvA) oscillations spectra exhibit frequency
combinations typical of such a network. Even though some of the
observed magnetoresistance oscillations cannot be interpreted on
the basis of neither conventional Shubnikov-de Haas oscillations
nor quantum interference, the temperature and magnetic field (both
orientation and magnitude) dependence of all the Fourier
components of the dHvA spectra can be consistently accounted for
by the dHvA effect on the basis of the Lisfhitz-Kosevich formula.
This behaviour is at variance with that currently reported for
compounds illustrating the linear chain of coupled orbits model.
\end{abstract}

\begin{figure}
\centering
\resizebox{\textwidth}{!}{\includegraphics*{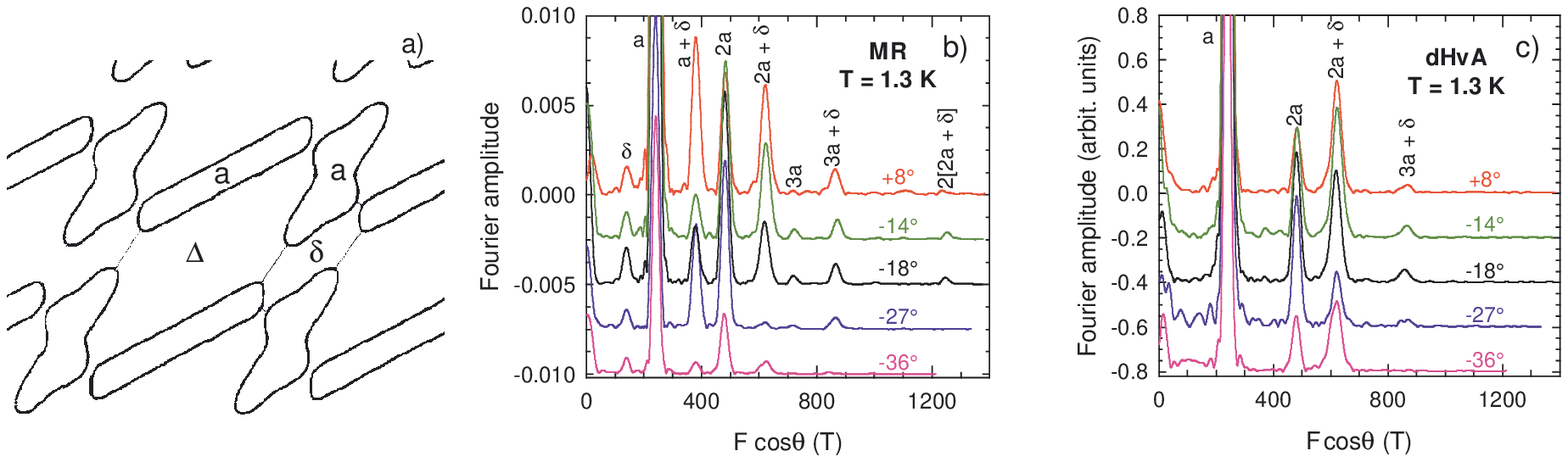}}
\caption{(a) Fermi surface of
(ET)$_8$Hg$_4$Cl$_{12}$(C$_6$H$_5$Cl)$_2$ according to band
structure calculations of Vieros et al. \cite{Vi94}. In addition
to the compensated elongated electron and hole closed orbits
\emph{a}, the $\delta$ and $\Delta$ pieces are indicated. Fourier
analysis of (b) magnetoresistance and (c) de Haas-van Alphen
oscillations for various directions of the magnetic field at 1.3
K.} \label{TF(angle)+SF}
\end{figure}

Frequency combinations observed in magnetic oscillations spectra
of multiband quasi two-dimensional (2D) metals have been
extensively studied both from theoretical and experimental
viewpoints \cite{Ka04}. Nevertheless, the physical origin of some
of the observed Fourier components, the so called 'forbidden
frequencies', remain still unclear. In addition to quantum
interference (QI) that can be invoked in the case of
magnetoresistance (MR) data, these frequencies can be attributed
to both the oscillation of the chemical potential\cite{Ha96, Ki02}
and the field-dependent magnetic breakthrough (MB)-induced Landau
level broadening\cite{brooks}. However, a quantitative model that
involve these two latter contributions is still needed. The Fermi
surface (FS) of most of the compounds exhibiting such frequencies
corresponds to the linear chain of coupled orbits model. This
model was introduced by Pippard in the early sixties in order to
compute MB-induced de Haas-van Alphen (dHvA) oscillation spectrum
in a one-dimensional network of electronic orbits\cite{Pi62}. More
recently, the 2D network of compensated electron and hole orbits
without any non-quantized electron reservoir, realized by the FS
of the isostructural organic metals
(ET)$_8$Hg$_4$Cl$_{12}$(C$_6$H$_5$X)$_2$ where ET stands for
bis(ethylenedithio)tetrathiafulvalene and X = Cl, Br (see Fig.
\ref{TF(angle)+SF}a and Ref. \cite{Vi94}), has been considered
\cite{Ly95, Pr02, Vi03}. Contrary to the $\beta$-(ET)$_2$IBr$_2$
salt whose oscillatory spectra exhibit either additional slow
oscillations or beatings due to a significantly warped FS
\cite{Ka02}, the above compounds are strongly 2D \cite{Ly95}. The
MB orbits and QI paths liable to be involved in the oscillatory
spectra are described in Refs. \cite{Pr02, Vi03}. The Fourier
spectra of the oscillatory MR have been analyzed on the basis of
the Falicov and Stachowiak model\cite{Fa66} which assumes that the
effective mass of a given MB orbit is the sum of the effective
masses of each of its components. In agreement with this model,
the effective mass of, e. g., the MB-induced 2\emph{a} + $\delta$
orbit observed in MR experiments is about twice that of the
compensated 'basic' orbits \emph{a}. In addition to these
Shubnikov-de Haas (SdH) orbits, some of the observed Fourier
components could be attributed to QI linked to e. g. \emph{a} +
$\delta$ and \emph{b} = 2\emph{a} + $\delta$ + $\Delta$
interferometers. Nevertheless, in view of the measured values of
the MB field and of the relevant effective masses, the field
dependence of the oscillation amplitude of \emph{a} + $\delta$ is
not in agreement with such a picture. Moreover, other components
such as $\delta$ which has a very low effective mass (see Table
\ref{table}) cannot be understood on the basis of neither SdH
orbits nor QI. Previous attempts at measuring the oscillatory
magnetization which, as a thermodynamic parameter, should be
insensitive to QI, revealed that the effective mass linked to SdH
and dHvA oscillations of the closed orbits \emph{a} and 2\emph{a}
+ $\delta$, respectively, is the same within the error
bars\cite{Vi03,Au04}. However, the signal-to-noise ratio was too
weak in order to obtain reliable results for other Fourier
components.

In this letter, we present simultaneous measurements of MR and
dHvA oscillations of (ET)$_8$Hg$_4$Cl$_{12}$(C$_6$H$_5$Br)$_2$ for
various directions of the magnetic field up to 28 T. Even though
'forbidden frequencies' are evidenced in MR oscillations spectra,
it is demonstrated that, at variance with compounds whose FS
illustrates the linear chain model, the behaviour of the frequency
combinations observed in dHvA spectra of this network of
compensated orbits can be consistently accounted for by the
Falicov and Stachowiak model on the basis of the Lifshits-Kosevich
(LK) formula.

\begin{figure}                                                  
\threefigures[scale=0.25]{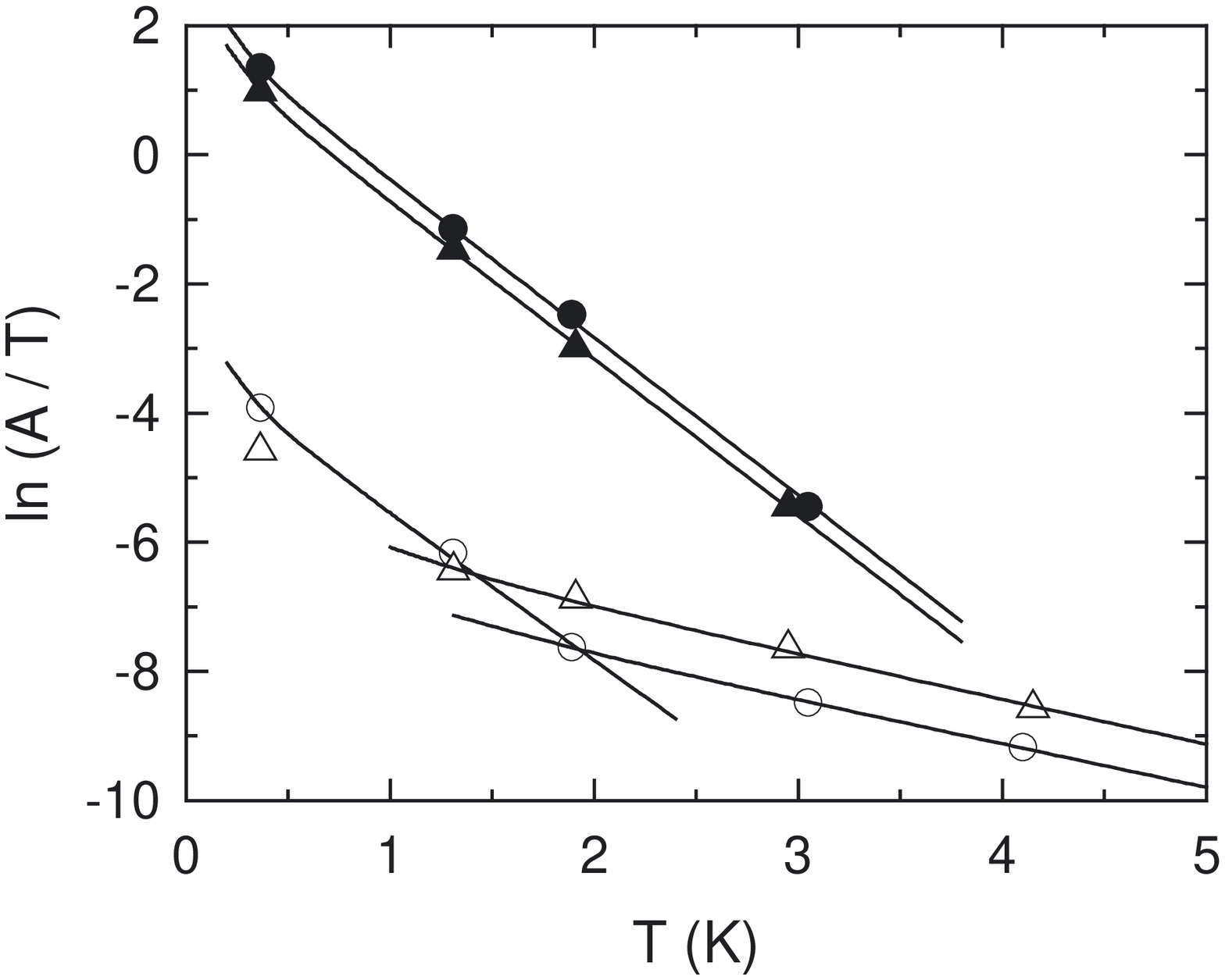}{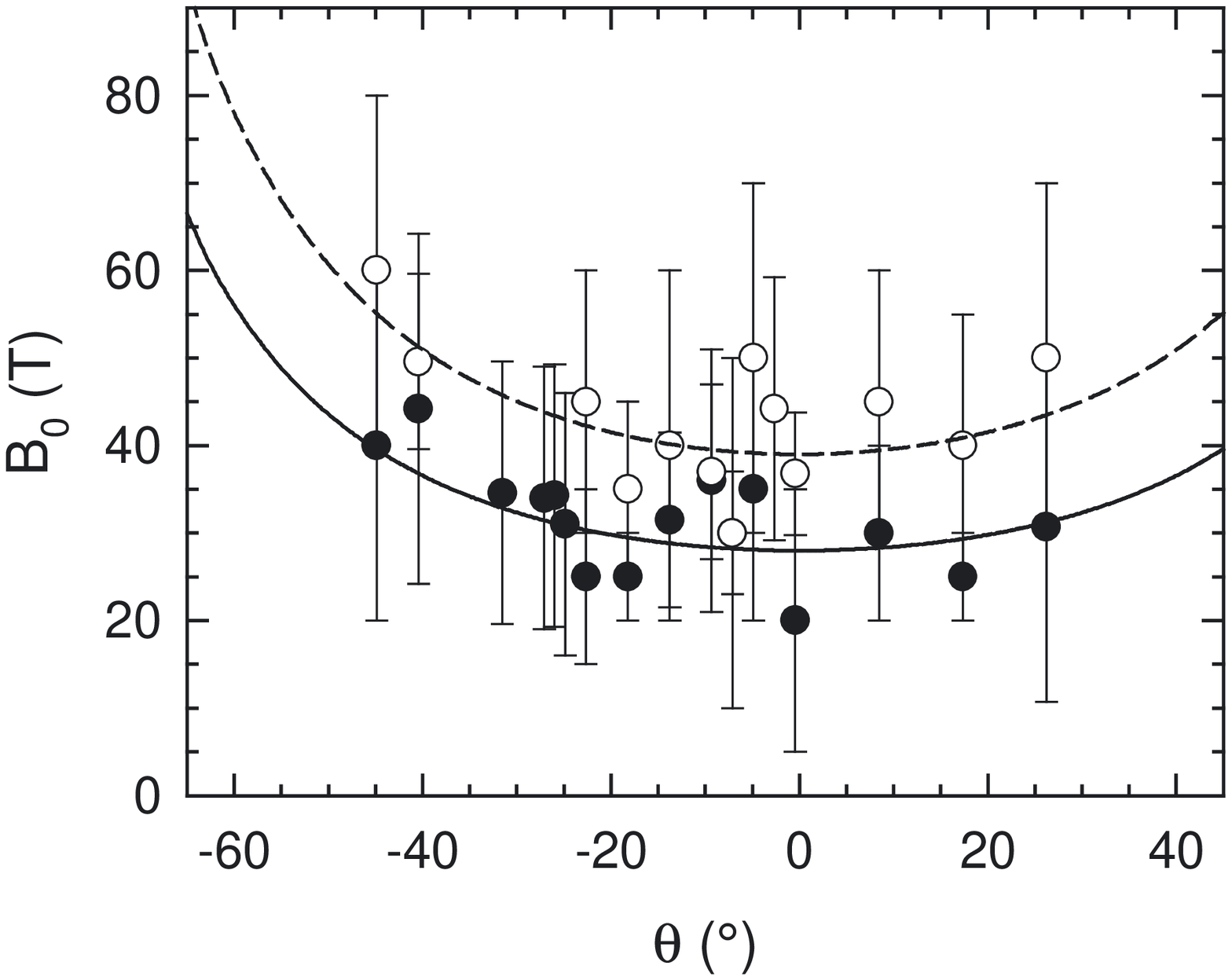}{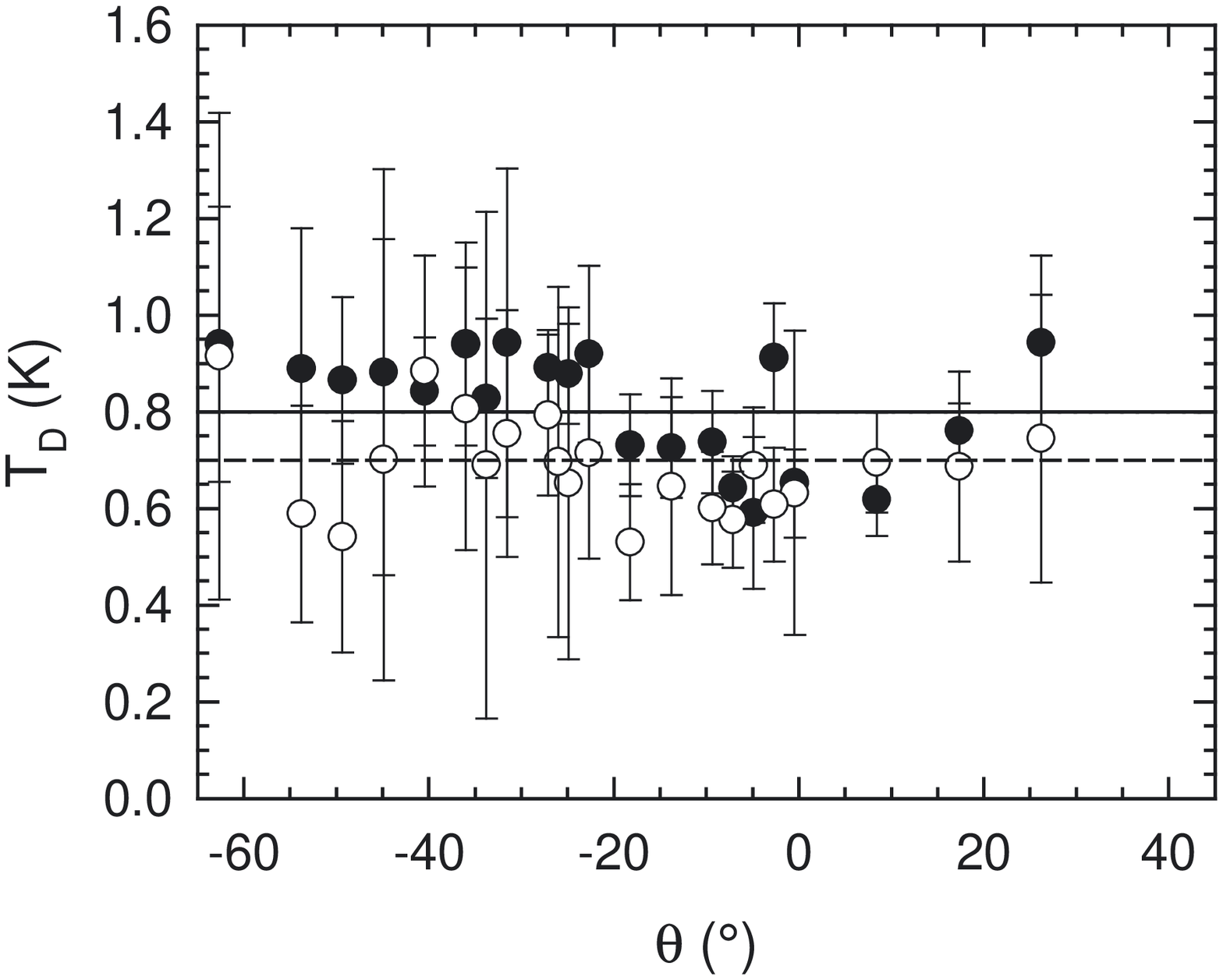}
\caption{Temperature dependence of the Fourier amplitude (A) of
magnetoresistance (MR, open symbols) and de Haas-van Alphen (dHvA,
solid symbols) data for the 3\emph{a} + $\delta$ component at a
mean value of the magnetic field $\overline{B}$ = 21.4 T. The
angle between the field direction and the normal to the conducting
plane is 17$^\circ$ and -27$^\circ$, for circles and triangles,
respectively. Solid lines are best fits of Eqs. \ref{LK_SdH} and
\ref{LK_dHvA} to MR and dHvA data, respectively.}
\label{lnAsurT_3a+d} \caption{Mean magnetic breakdown field values
(B$_0$) deduced from the field dependence of the amplitude of the
\emph{a} oscillations at 1.3 K for various field directions.
Closed and open symbols stand for dHvA and Shubnikov-de Haas (SdH)
data, respectively. Solid and dashed line are best fits to the
dHvA and SdH data, respectively, assuming an orbital behaviour.}
\label{B0} \caption{Same as Fig. \ref{B0} for the Dingle
temperature. Solid and dashed line corresponds to the mean value
of dHvA (T$_D$ = 0.8 K) and SdH (T$_D$ = 0.7 K) data,
respectively.}
  \label{TD}
\end{figure}

The studied crystal is a platelet with approximate dimensions
(1$\times$1$\times$0.1)~mm$^3$, the largest sides being parallel
to the conducting \emph{bc}-plane. MR and magnetic torque
measurements were performed simultaneously up to 28 T at the M10
resistive magnet of the Grenoble High Magnetic Field Laboratory. A
one axis rotating sample holder allowed to change the direction of
the magnetic field with respect to the conducting plane. In the
following, $\theta$ is the angle between the magnetic field
direction and the $a^*$ axis. Experiments were performed in the
temperature range from 0.36 K to 4.2 K for $\theta$ = -27$^\circ$,
-18$^\circ$, 8$^\circ$ and 17$^\circ$ and at 1.3 K for -62$^\circ$
$\leq \theta \leq$ 26$^\circ$ (the sign of $\theta$ is arbitrary).
Electrical contacts were made to the crystal for MR measurements
using annealed gold wires of 10 $\mu$m in diameter glued with
graphite paste. Alternating current (10 $\mu$A, 10 Hz) was
injected parallel to the \emph{a}* direction (interlayer
configuration). A lock-in amplifier was used to detect the signal
across the potential leads. Magnetic torque was measured by the
cantilever method\cite{Sh03}. Its oscillatory part is given by
$\tau$ =
-(1/F)M$_{\parallel}$(${\partial}$F/${\partial}$$\theta$)BV
\cite{Wo96, Sh03, Sh84} where M$_{//}$ is the component of the
oscillatory magnetization parallel to B and V is the crystal
volume. For a 2D FS, the oscillation frequency varies as F =
F($\theta$ = 0)/cos($\theta$) and ${\tau}$ is therefore
proportional to M$_{//}$B$\tan \theta$.

In the framework of the 2D LK model, the measured Fourier
amplitudes (A$_i$) of the various components of the normalized
oscillatory MR ($R_{osc}/R_{bg}-1$, where $R_{osc}$ and $R_{bg}$
is the oscillatory and background resistance, respectively) and of
the torque are respectively given by \cite{Sh84}:
\begin{equation}
\label{LK_SdH}
A_i(MR) \propto R_{Ti}R_{Di}R_{MBi}R_{Si}
\end{equation}
and
\begin{equation}
\label{LK_dHvA}
A_i(torque) \propto BR_{Ti}R_{Di}R_{MBi}R_{Si}\tan\theta,
\end{equation}

For a 2D FS, the measured effective mass (expressed in m$_e$ units
in the following) depends on the field direction as m${_i^*}$ =
m${_i^*}$($\theta$ = 0)/cos($\theta$). This angle dependence has
been checked in Refs. \cite{Vi03, Au04} for the presently studied
compound. The thermal and Dingle damping factors are therefore
respectively given by R$_{Ti}$ = $\alpha$Tm${_i^*}$($\theta$ =
0)/Bcos($\theta$)sinh[$\alpha$Tm${_i^*}$($\theta$ = 0)/B
cos($\theta$)] and R$_{Di}$ = exp[-$\alpha$T$_D$m${_i^*}$($\theta$
= 0)/B cos($\theta$)], where $\alpha$ =
2$\pi^2$k$_B$m$_e$/e$\hbar$ ($\simeq$ 14.69 T/K) and T$_D$ is the
Dingle temperature. The spin and MB damping factors are given by:

\begin{equation}
\label{RS} R_{Si} = \mid\cos(\pi \mu/\cos\theta)\mid
\end{equation}

and

\begin{equation}
\label{RMB}
 R_{MBi} = \prod_{g=1,2}p_g^{t_{g}}q_g^{b_{g}},
\end{equation}

respectively, where $\mu$ = g$^*$m${_i^*}$($\theta$ = 0)/2 and
g$^*$ is the effective Land\'{e} factor. The index g stands for
the 2 different gaps between electron and hole orbits (see Fig.
\ref{TF(angle)+SF}a and Ref. \cite{Vi03}). Integers t$_{g}$ and
b$_{g}$ are respectively the number of tunnelling and Bragg
reflections encountered along the path of the quasiparticle. The
tunnel and Bragg reflection probability amplitudes are given by
p$_g^2$ = exp(-B$_g$/B), where B$_g$ is the MB field, and p$_g^2$
+ q$_g^2$ = 1. As discussed in Ref.~\cite{Vi03}, only a mean value
B$_0$ of the two MB fields can be derived from experimental data.
Eq. \ref{RMB} then reduces to R$_{MBi}$ = p$^{t}$q$^{b}$ where
p$^2$ = exp(-B$_0$/B). Calculations of R$_{MBi}$ and effective
masses for various orbits are given in Ref. \cite{Vi03}.

In addition to the frequency linked to the 'basic' orbits
\emph{a}, Fourier analysis of the MR oscillations reveals
combinations frequencies involving the $\delta$ piece of the FS,
in good agreement with data from Ref. \cite{Vi03} (see Fig.
\ref{TF(angle)+SF}b). Nevertheless, since the maximum field
reached in the experiments is lower than in Ref. \cite{Vi03}, the
high frequency components, in particular those involving the
$\Delta$ piece of the FS cannot be observed. As it is the case for
MR data, dHvA spectra also present frequency combinations.
Nevertheless, $\delta$ and \emph{a} + $\delta$ components are
clearly absent in Fig. \ref{TF(angle)+SF}c. Except for few MR
components at high field, as developed later on, the temperature
dependence of the observed oscillations is satisfactorily
accounted for by the LK formalism. The deduced values of the
effective masses are reported in Table \ref{table}. A good
agreement between MR and dHvA data is observed for \emph{a} and
2\emph{a} + $\delta$ closed orbits. Oppositely, the behaviour of
the 3\emph{a} + $\delta$ component calls for some comments.
Indeed, a change of the effective mass deduced from MR data has
been reported in Ref.~\cite{Vi03}, in agreement with data in Fig.
\ref{lnAsurT_3a+d}. Generally speaking, the effective mass deduced
from low temperature data is in agreement with a MB-induced orbit
while the high temperature value, which is about 3 times lower, is
consistent with QI. In contrast, no change of the effective mass
is observed in dHvA data which otherwise are in agreement with a
MB-induced orbit.  In summary, at variance with MR data, the
effective mass values deduced from all the Fourier components of
the torque data can be interpreted, in the framework of the
Falicov and Stachowiak model, on the basis of dHvA effect linked
to, eventually MB-induced, closed orbits.

\begin{table}                                                          
\caption{\label{table}Effective mass [m$^*$($\theta$ = 0)] and
$\mu$ parameter ($\mu$ = g$^*$m${_i^*}$($\theta$ = 0)/2) deduced
from temperature and field direction dependence, respectively, of
the Fourier components indicated in the left column. }
\begin{tabular}{c|cccc|cc}
&\multicolumn{4}{c|}{m$^*$($\theta$ = 0)}&\multicolumn{2}{c}{$\mu$}\\
&MR\cite{Vi03}&dHvA\cite{Au04}&MR&dHvA&MR&dHvA\\
\hline
$\delta$&0.45$\pm$0.10&&0.3$\pm$0.1&&&\\
a&1.17$\pm$0.13&1.15$\pm$0.12&1.23$\pm$0.04&1.23$\pm$0.04&1.28$\pm$0.04&1.25$\pm$0.03\\
a + $\delta$&1.00$\pm$0.07&&1.06$\pm$0.07&&&\\
2a + $\delta$&2.10$\pm$0.16&2.10$\pm$0.25&2.15$\pm$0.15&2.15$\pm$0.13&2.2$\pm$0.1&2.15$\pm$0.15\\
3a + $\delta$ (high T)&0.72$\pm$0.06&&0.9$\pm$0.1&&\\
3a + $\delta$ (low T)&2.95$\pm$0.20&&3.1$\pm$0.3&\raisebox{1.7ex}[1ex]{3.1$\pm$0.3}&&\raisebox{1.7ex}[1ex]{2.75$\pm$0.20}\\

\end{tabular}
\end{table}

\begin{figure} [h]
\resizebox{\textwidth}{!}{\includegraphics*{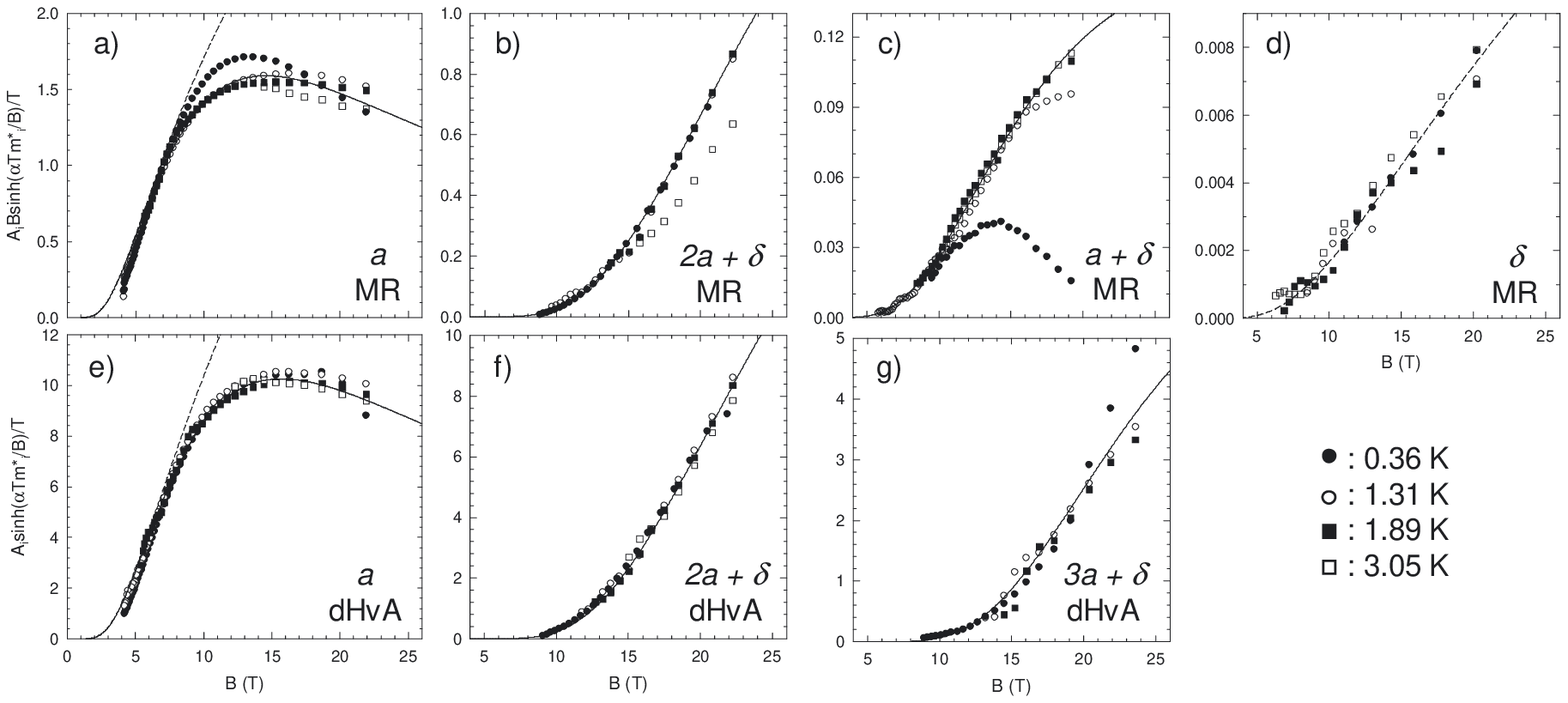}}                
\caption{Magnetic field dependence, for $\theta$ = 17$^{\circ}$,
of the Fourier amplitude of (a) to (d) magnetoresistance (MR) and
(e) to (g) de Haas-van Alphen (dHvA) components normalized by the
relevant thermal damping factor R$_{Ti}$ for i = \emph{$\delta$},
\emph{a}, \emph{a} + \emph{$\delta$}, 2\emph{a} + \emph{$\delta$}
and 3\emph{a} + \emph{$\delta$}. Solid lines are best fits of Eqs.
\ref{LK_SdH} and \ref{LK_dHvA} to MR and dHvA data, respectively,
with B$_0$ = 25 T. Dashed lines are obtained with R$_{MBi}$ = 1.
The Dingle temperature is in the range 0.6 K to 0.7 K, except for
\emph{$\delta$} which cannot be accounted for by a closed orbit
and \emph{a} + \emph{$\delta$} for which the solid line
corresponds to a negative value. } \label{A(B)}
\end{figure}

This can further be checked by considering the field dependence of
the Fourier components. A quantitative analysis of these data
requires first the determination of the MB field B$_0$. Since the
\emph{a} orbit only involves Bragg reflections, the most
pronounced damping of the oscillation amplitude as the magnetic
field increases is observed for this component. This feature
allows us to determine B$_0$, as well as T$_D$, from the field
dependence of the amplitude of this oscillation series. The angle
dependence of the MB field deduced from the analysis of the data
collected at 1.3 K in the range -62$^\circ$ $\leq \theta \leq$
26$^\circ$ is compatible with an orbital behaviour and yields
B$_0$($\theta$ = 0) = (28 $\pm$ 9) T and (39 $\pm$ 10) T, for dHvA
and SdH oscillations, respectively (see Fig. \ref{B0}). Besides,
the Dingle temperature reported in Fig. \ref{TD} can be regarded
as temperature-independent: T$_D$ = (0.8 $\pm$ 0.2) K and T$_D$ =
(0.7 $\pm$ 0.2) K for dHvA and SdH data, respectively. According
to the LK formula, the temperature-independent part of the
amplitude is given by A$_i$/R$_{Ti}$. As an example, data for
$\theta$ = 17$^\circ$ are given in Fig. \ref{A(B)}. In agreement
with the LK model, all the dHvA data related to the \emph{a}
component lie on the same curve (see Fig. \ref{A(B)}e) and are
accounted for by B$_0$ = 25 T and T$_D$ = 0.7 K. The same value of
effective mass as for dHvA data is used for the analysis of MR
data in Fig. \ref{A(B)}a. Data at low field yield T$_D$ = 0.6 K
which is in good agreement with data of  Fig. \ref{TD} and with
dHvA data of Fig. \ref{A(B)}e. However, some discrepancies can be
observed at fields higher than a few teslas, in particular at the
lowest temperature. This feature puts some doubts on the value of
B$_0$ extracted from MR data, which otherwise is in good agreement
with data from Ref. \cite{Vi03} [B$_0$ = (35 $\pm$ 7) T].
Regarding the MB-induced 2\emph{a} + \emph{$\delta$} orbit, both
MR and dHvA data are consistent with each other and with the data
for the \emph{a} component. Namely, assuming B$_0$ = 25 T, the
solid line in Figs. \ref{A(B)}b and f correspond to T$_D$ = 0.7 K,
in good agreement with data for the \emph{a} oscillations.
Analogous result is obtained for dHvA data related to the
3\emph{a} + \emph{$\delta$} component (T$_D$ = 0.7 K, see Fig.
\ref{A(B)}g). Oppositely, the \emph{a} + \emph{$\delta$} component
observed in MR, which has been attributed to QI, exhibits
significant discrepancies with the LK formalism at low temperature
and high field (see Fig. \ref{A(B)}c). In addition, the Dingle
temperature value deduced from the solid line in the figure is
negative (T$_D$ = - 0.5 K), as it is the case for the data
reported in Ref. \cite{Vi03}. Finally, MR data for the
\emph{$\delta$} component (see Fig. \ref{A(B)}d) follow the LK
behaviour even though this component cannot be interpreted on the
basis of the Falicov and Stachowiak model. Recall that MR data for
3\emph{a} + \emph{$\delta$} cannot be considered due to the
observed change of the effective mass as the temperature varies.

\begin{figure}
\resizebox{\textwidth}{!}{\includegraphics*{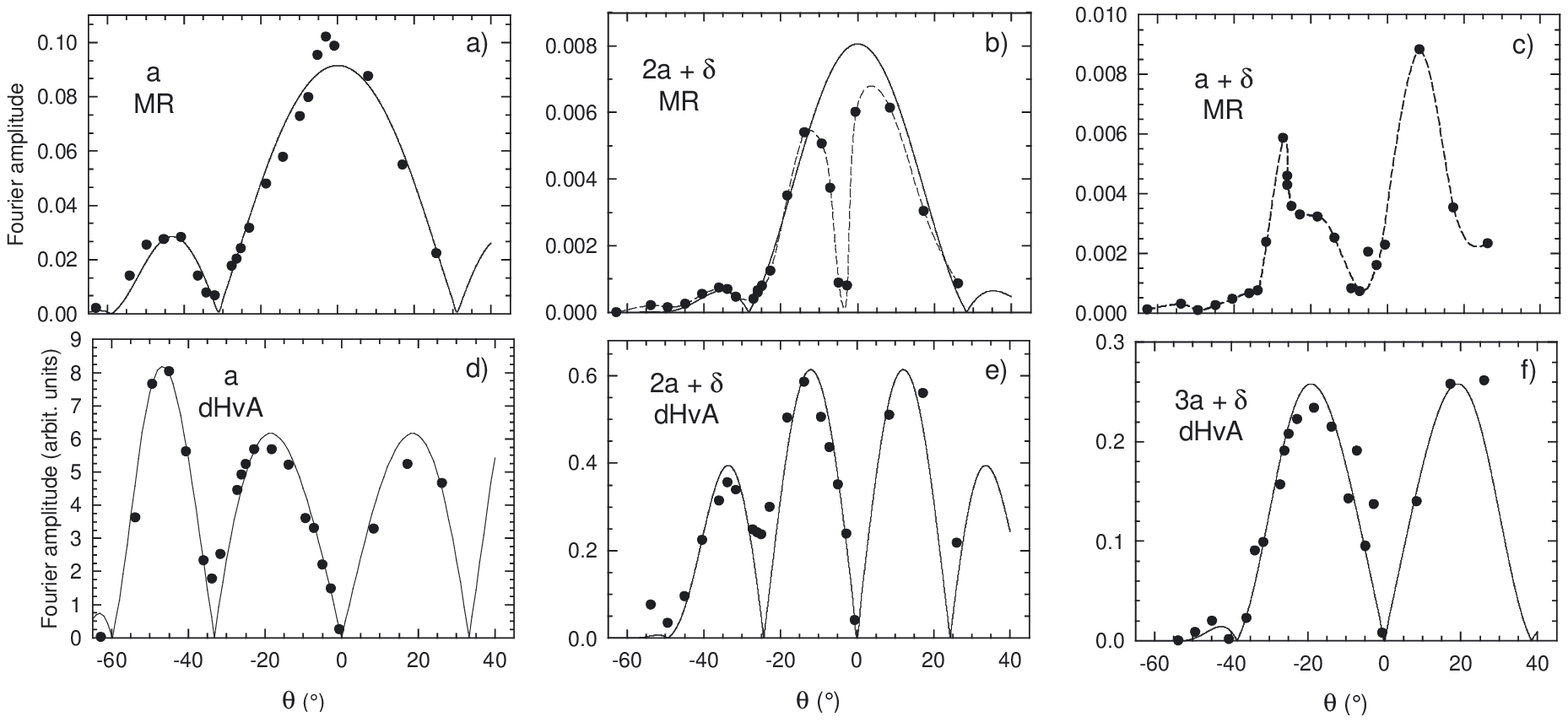}}                
\caption{Angle ($\theta$) dependence of the (a, b, c)
magnetoresistance (MR) and (d, e, f) de Haas-van Alphen (dHvA)
Fourier amplitudes for \emph{a}, \emph{a} + \emph{$\delta$},
2\emph{a} + \emph{$\delta$} and 3\emph{a} + \emph{$\delta$}
components at a temperature of 1.3 K and a mean value of the
magnetic field $\overline{B}$ = 14.7 T. Solid lines are best fits
of Eqs. \ref{LK_SdH} and \ref{LK_dHvA} to MR and dHvA data,
respectively. Dashed lines are guides to the eye.} \label{spin0}
\end{figure}

Lastly, let us consider the field direction dependence of the
oscillations amplitude for which the spin damping factor (see Eq.
\ref{RS}) plays an important part. A good agreement with the LK
model is observed for both MR and dHvA data related to the
\emph{a} oscillations (see Fig. \ref{spin0}). In addition, the
deduced values of $\mu$ are very close to m$^*$ (see Table
\ref{table}) which is in agreement with a Fermi liquid, as
suggested in Ref. \cite{Ly04}. The same feature is observed for
dHvA data of the 2\emph{a} + \emph{$\delta$} and 3\emph{a} +
\emph{$\delta$} components. MR data for the MB orbit 2\emph{a} +
\emph{$\delta$} evidence a strong damping at $\theta \approx$
-5$^\circ$. Analogous damping of the MB orbit have also been
observed, although at $\theta$ = 0$^\circ$, in MR data of
$\kappa$-(ET)$_2$I$_3$ \cite{Ba96, Ha98}. Some damping of the SdH
oscillations linked to the 'basic' orbit of
$\beta$''-(ET)$_2$SF$_5$CH$_2$CF$_2$SO$_3$ is also reported in the
high field and low temperature ranges\cite{Wo99}. This behaviour
which is not observed for dHvA oscillations \cite{Ba96, Ha98,
Wo99} has still not received a clear explanation. In spite of the
damping observed in Fig. \ref{spin0}b, the data away from $\theta
\approx$ -5$^\circ$ is still in agreement with a $\mu$ value close
to m$^*$. In contrast, the amplitude of the \emph{a} +
\emph{$\delta$} component, which is not symmetric with respect to
$\theta$, cannot be accounted for by the LK formula. As it is the
case of 2\emph{a} + \emph{$\delta$}, it exhibits a minima around
$\theta \approx$ -5$^\circ$.

In summary, all the observed torque components can be consistently
interpreted on the basis of dHvA closed orbits. This is at
variance with MR oscillations linked to (i) SdH closed orbits
\emph{a} and 2\emph{a} + \emph{$\delta$} which exhibit
discrepancies at high field and at $\theta \approx$ -5$^\circ$,
respectively, and (ii) other components such as \emph{a} +
\emph{$\delta$} for which the field dependence is too weak in
order to account for a positive T$_D$ even though its effective
mass is consistent with QI. Departures from the LK behaviour have
already been reported for many clean crystals of quasi-2D organic
metals. Therefore, the observed MR behaviour is not surprising.
Regarding dHvA data, 'forbidden' frequency combinations with
significant amplitude are observed in compounds illustrating the
linear chain model. This is at variance with the presently
observed behaviour. As suggested in Ref. \cite{Fo05}, this
discrepancy could be due, for a part, to the nature of the
considered orbits network for which the chemical potential might
be pinned between electron and hole Landau levels. This problem,
that actually remains to be solved, is of importance since the FS
of many organic metals based on the ET molecule or one of its
derivative constitutes a network of compensated electron and hole
orbits as well.

\acknowledgments  We wish to acknowledge J. Y. Fortin and G.
Rikken for interesting discussions. This work was supported by the
CNRS-RAS cooperation agreement $\#$ 16390. RBL thanks RFBR
03-02-16606 for support.


\begin{thebibliography}{0}

\bibitem [1]{Ka04}
 For a review, see e. g.
 \Name{Kartsovnik M.}
 \REVIEW{Chem. Rev.} {104} {2004} {5737} and references therein.

\bibitem [2]{Ha96}
 \Name{Harrison N., Caulfield J., Singleton J., Reinders P. H. P., Herlach F.,
 W. Hayes, M. Kurmoo and P. Day}
 \REVIEW{J. Phys.: Condens. Matter} {8} {1996} {5415}.

\bibitem[3]{Ki02}
 \Name{Kishigi K. and Hasegawa Y.}
 \REVIEW{Phys. Rev. B} {65} {2002} {205405}.

\bibitem[4]{brooks}
 \Name{Sandhu P. S., Kim J. H. and Brooks J. S.}
 \REVIEW{Phys. Rev. B} {56} {1997} {11566}.
 \Name{Fortin J. Y. and Ziman T.}
 \REVIEW{Phys. Rev. Lett.} {80} {1998} {3117}.
 \Name{Kim J. H., Han S. Y. and Brooks J. S.}
 \REVIEW{Phys. Rev. B} {60} {1999} {3213}.
 \Name{Han S. Y., Brooks J. S. and Kim J. H.}
 \REVIEW{Phys. Rev. Lett.}{85} {2000} {1500}.
 \Name{Gvozdikov V. M., Pershin Yu V., Steep E., Jansen A. G. M. and Wyder P.}
 \REVIEW{Phys. Rev. B} {65} {2002} {165102}.

\bibitem [5]{Pi62}
 \Name{Pippard A. B.}
 \REVIEW{Proc. Roy. Soc. (London)} {A270} {1962} {1}.

\bibitem[6]{Vi94}
 \Name{Vieros L. F. and Canadell E.}
 \REVIEW{J. Phys. (France) I} {4} {1994} {939}.

\bibitem[7]{Ly95}
 \Name{Lyubovski\u{i} R. B., Pesotski\u{i} S. I., Gilevski\u{i} A. and Lyubovskaya R. N.}
 \REVIEW{JETP} {80} {1995} {1063} [\REVIEW{Zh. \'{E}ksp. Teor.
 Fiz.} {107} {1995} {1698}].


\bibitem[8]{Pr02} \Name{Proust C., Audouard A., Brossard L., Pesotski\u{i} S. I., Lyubovski\u{i} R. B.
 and Lyubovskaya R. N.}
 \REVIEW{Phys. Rev. B} {65} {2002} {155106}.

\bibitem[9]{Vi03}
 \Name{Vignolles D., Audouard A., Brossard L., Pesotski\u{i} S. I.,
B. Lyubovski\u{i} R. and Lyubovskaya R. N.}
 \REVIEW{Eur. Phys. J. B} {31}{2003} {53}.

\bibitem [10]{Ka02}
 \Name{M. V. Kartsovnik, P. D. Grigoriev, W. Biberacher, N. D. Kushch and P. Wyder}
 \REVIEW{Phys. Rev. Lett.} {89} {2002} {126802}.

\bibitem [11]{Fa66}
 \Name{Falicov L. M. and Stachowiak H.}
 \REVIEW{Phys. Rev.} {147} {1966} {505}.

\bibitem[12]{Au04}
 \Name{Audouard A., Vignolles D., Proust C., Brossard L., Nardone M., Haanappel E., Pesotski\u{i} S. I.,
B. Lyubovski\u{i} R. and Lyubovskaya R. N.}
 \REVIEW{Physica B} {346-347} {2004} {377}.

\bibitem[13]{Sh03} \Name{Sheikin I., Gr\"{o}ger A., Raymond S., Jaccard D., Aoki D., Harima H. and Flouquet J.}
 \REVIEW{Phys. Rev. B} {67} {2003} {94420}.

\bibitem[14]{Sh84}
  \Name{Shoenberg D.}  
  \Book{Magnetic Oscillations in Metals}
  \Publ{Cambridge Univ. Press, Cambridge}
  \Year{1984}.

 \bibitem[15]{Wo96}
  \Name{Wosnitza J.}  
  \Book{Fermi Surfaces of Low-dimensional Organic Metals and Superconductors}
  \Publ{Springer-Verlag Berlin Heidelberg}
  \Year{1996}.

\bibitem[16]{Ly04}
 \Name{Lyubovski\u{i} R. B., Pesotski\u{i} S. I., Nizhankovski\u{i} V. I., Biberacher W. and Lyubovskaya R. N.}
 \REVIEW{JETP} {98} {2004} {1037} [\REVIEW{Zh. \'{E}ksp. Teor.
 Fiz.} {125} {2004} {1184}].

\bibitem[17]{Ba96}
 \Name{Balthes E., Schweitzer D., Heinen I., Strunz W., Biberacher W., Jansen A. G. M. and Steep E.}
 \REVIEW{Z. Phys. B} {99} {1996} {163}.

\bibitem [18]{Ha98}
 \Name{Harrison N., Mielke C. N., Rickel D. G., Wosnitza J., Qualls J. S., Brooks J. S., Balthes E.,
 Schweitzer D., Heinen I. and Strunz W.}
 \REVIEW{Phys. Rev. B} {58} {1998} {10248}.

\bibitem[19]{Wo99}
  \Name{Wosnitza J., Wanka S., Qualls J. S., Mielke C. H., Harrison N., Schlueter J. A., Williams J. M.,
  Nixon P. G., Winter R. W. and Gard G. L.}
  \REVIEW{Synth. Met.}{103}{1999}{2000}.

\bibitem [20]{Fo05}
 \Name{Fortin J. Y., Perez E. and Audouard A.}
 \REVIEW{Phys. Rev. B} {71} {2005} {155101}.





\end{thebibliography}
\end{document}